\documentclass[12pt,twoside]{article}
\usepackage{amssymb}
\usepackage{amsmath}
\usepackage{latexsym}
\usepackage{longtable}
\usepackage{epsfig}
\usepackage{graphicx,bbm,psfrag}
\graphicspath{{images/}}
\usepackage{cite}
\usepackage{authblk}

\begin{document}
\begin{titlepage}
\vspace*{2cm}
\begin{center}
\Large{\textbf{Ballistic Space-Time Correlators of the Classical Toda Lattice}}\bigskip\\
{\large Herbert Spohn}
 \end{center}
 \begin{center}
Zentrum Mathematik and Physik Department, TUM,\\
Boltzmannstra{\ss}e 3, 85747 Garching, Germany,
 \tt{spohn@tum.de}
 \end{center}
\vspace{3cm}
\textbf{Abstract}: The Toda lattice is an integrable system and its natural space-time stationary states are the 
generalized Gibbs ensembles (GGE). Of particular physical interest are then the space-time correlations of the conserved fields.
To leading order they scale ballistically. We report on the exact solution of the respective generalized hydrodynamic equations
linearized around a GGE as background state. Thereby we obtain a concise formula for the family of scaling functions. 
\vspace*{2cm}
\begin{flushright} 25.11.2019
\end{flushright}
\end{titlepage}
\section{Introduction}
\label{sec1}
\setcounter{equation}{0} 
With the discovery of the Toda lattice as an integrable classical field theory an immediate hope was to be able to compute time-correlations in thermal equilibrium. During the 1980ies considerable efforts were invested \cite{SchS80,D81,Sch83,TM83,T84,Opper,TI86,GM88,CSTV93,JE00}. But mostly a better understanding of static properties were accomplished \cite{TM83, T84,Opper,TI86}. 
A few at the time pioneering molecular dynamics simulations became available \cite{SchS80, Sch83}. Theoretically  low order continued fraction expansions were tried, also expansions close to the two solvable limits, namely harmonic chain and hard rods \cite{D81,CSTV93}. More recently, with improved computational power, molecular dynamics simulations with approximately 1000 particles have been performed \cite{KD16}.
Time-displaced correlations as stretch-stretch (the same as free volume) and energy-energy are measured with high precision, which requires not only numerically solving  a system of Newton's equations of motion but also generating roughly $10^7$ samples  so to keep the noise level small. As one  main conclusion from these studies,  the mentioned correlations scale ballistically over the accessible time span, except for very early times.  Thus the challenging goal is to  predict, resp. to compute, the scaling function for a specified correlation. Another  important advance happened on the theoretical side. Starting with integrable quantum field theories the formalism of generalized hydrodynamics (GHD) has been developed \cite{CDY16,PNCBF17}, which in fact also applies to classical counterparts, in particular to the Toda lattice \cite{D19,BCM19}. In the context of time correlations only the equations linearized around thermal equilibrium are of relevance.
Some preliminary results are reported in \cite{D19,S19}. In this contribution we provide the exact solution of linearized hydrodynamics. 
For quantitative results one still has to numerically
determine the density of states (DOS) of the Lax matrix and solve the linear integral equation linked to the dressing transformation. To fully exploit the linear structure all conserved quantities have to be treated on equal footing. Also thermal equilibrium is just one particular case
of a generalized Gibbs ensemble (GGE).

The connection between linear response and microscopic  correlation functions is standard wisdom. In \cite{D19b} this theme is studied in detail for a 
general class of 
integrable systems governed by GHD. For the one-dimensional sinh-Gordon model molecular dynamics is compared with the predictions from linearized GHD
\cite{BDWY18}.
While our arguments follow a somewhat different route, our final result on time-correlations of the conserved fields is in complete agreement with the general theory developed in \cite{D19b}.

For the Toda lattice, thermal, more generally GGE, average of the conserved fields is most concisely encoded through the DOS of the Lax matrix. But on top there is the conserved stretch which does not seem to be connected to the Lax matrix and plays a special role.
The stretch  current equals momentum, hence also conserved, which is not the case for all the other conserved fields. As a consequence generalized hydrodynamics consists of two equations, one for the stretch and one for the Lax DOS. The linearized equations then inherit a 
$2\times 2$ matrix structure (with operator entries) and one difficulty in the analysis is to properly  handle this feature. Such property appears also, for example, in the XXZ 
chain in which  case the exceptional role is played by the magnetization.  The insight from the Toda lattice might thus be helpful for other integrable models.

A basic claim of generalized hydrodynamics is to be valid over the entire parameter range, no exceptions. This claim is particularly intriguing for the Toda chain. In thermal equilibrium, for large $N$,
the length of the system behaves as $\nu N$ with some coefficient $\nu(P)$, which depends on the applied pressure $P$. $\nu$ can have either sign. 
With high probability, for small $P$ particles are ordered, while for negative  $P$ they are anti-ordered. But this means that there is a specific pressure $P_\mathrm{c}$ at  which $\nu(P_\mathrm{c}) = 0$ and the typical distance between particles is of order  $1/\sqrt{N}$ only. So the picture that 
the interactions can be broken up into small groups of particles, which are isolated from the rest of the system and move from an incoming configuration to an outgoing one, becomes questionable.  On the Euler level the propagation speed of a perturbation is proportional to $\nu^{-1}$, thus formally diverges.
It is an open problem to understand the dynamical behavior close to $\nu =0$.

Our predictions are valid for the ballistic scale and one may wonder how much microscopic information is still visible. We recall that  for a simple fluid the Euler equations have always the same structure. The particular fluid under consideration is encoded by the thermodynamic pressure as a function of density and internal energy.   
For an integrable many-body system, apparently the system specific character comes from the two-particle phase shift, whereas the mathematical structure 
persists over a large  variety of models both quantum and classical. For the Toda lattice the phase shift is given by 
$2\log|v-v'|$ in dependence on 
the incoming velocities $v,v'$. The entire set of equations of generalized hydrodynamics could be written down by substituting some other function of the momentum transfer. Of course, there is then 
no reason to have an underlying microscopic particle model, the only confirmed cases being the Toda chain and hard rods, for which the phase shift equals the hard rod length $a$ independently of $v,v'$.

In Sect. \ref{sec2} we discuss some background material on the Toda chain. A summary on the generalized free energy is shifted to Appendix \ref{sec8}.
In Sect. \ref{sec3} we work out a convenient form of the static GGE charge-charge and charge-current correlations. This allows us to state then the exact solution of linearized hydrodynamics
with GGE random initial conditions. 
The required intermediate steps are reported in Sect. \ref{sec4}. There is an alternative way to obtain the static charge-charge correlator, which is related to Dyson's Brownian motion,
see Sect. \ref{sec5}. Hard rods turn out to be a useful guiding example, well covered in the literature \cite{BDS83,S91,BS97,DS17a}. To be closer to the Toda lattice, in Sect. \ref{sec6} we  study hard rods as coming from an anharmonic chain 
with an infinite hard core potential of core size $a$.  Thereby the structure of the corresponding generalized hydrodynamic equations resembles more closely the one of Toda.

\section{Conserved charges, currents, and normal form}
\label{sec2}
\setcounter{equation}{0} 
The Toda lattice is an anharmonic chain, for which the  interaction potential is specified to be exponential. The corresponding  hamiltonian is written as 
\begin{equation}\label{2.1}
H = \sum_{j\in\mathbb{Z}}\big( \tfrac{1}{2}p_j^2 + \mathrm{e}^{-r_j}\big),\quad r_j = q_{j+1} - q_j.
\end{equation}
Here $q_j,p_j$ is the position and momentum of the $j$-th particle.  The positional increments, $r_j$, are called stretches. We consider the 
infinitely extended lattice and indicate suitable finite number approximations whenever required. Then the equations of motion read
 \begin{equation}\label{2.2}
\frac{d}{dt}q_j = p_j, \qquad \frac{d}{dt}p_j =  \mathrm{e}^{-r_{j-1}}  -\mathrm{e}^{-r_j},\quad j \in \mathbb{Z},
\end{equation}
which is viewed as a discrete nonlinear wave equation.
Let us introduce the Flaschka  variables \cite{F74}, 
\begin{equation}\label{2.3}
a_j = \mathrm{e}^{-r_j/2},\quad b_j =  p_j.
\end{equation} 
In fact  the original Flaschka variables carry both a prefactor $\tfrac{1}{2}$ and, in principle, any prefactor could be used. However, only with our choice the generalized free energy has a particularly simple form, see Appendix \ref{sec8}. The Lax matrix  is  the tridiagonal real symmetric matrix with matrix elements
\begin{equation}\label{2.4}
L_{j,j} = b_j, \quad L_{j,j+1} = L_{j+1,j}= a_j. 
\end{equation} 
For the finite lattice $[1,....,N]$ the $N$ eigenvalues of $L$ are conserved. As functions on phase space they are non-local, their local version being 
$\mathrm{tr}[L^n]$, $n =1,2,...$ \cite{H74}. In generalized hydrodynamics such conserved fields are usually called charges or conserved charges
and it is convenient to follow this practice. The locally conserved charges of the Toda lattice have a strictly local density given by
\begin{equation}\label{2.5} 
 Q^{[n]}_j = (L^n)_{j,j},\quad n= 1,2,...,  
\end{equation}
with $j \in \mathbb{Z}$. In addition, the stretch is conserved with density
\begin{equation}\label{2.5a} 
 Q^{[0]}_j = r_j. 
\end{equation}
The respective  current densities are of the form
\begin{equation}\label{2.6} 
J^{[0]}_j = - Q^{[1]}_j , \qquad J^{[n]}_j = \tfrac{1}{2}(L^nL^\mathrm{off})_{j,j}.
\end{equation}
where $L^\mathrm{off}$ denotes the off-diagonal part of $L$ \cite{S19}.

For the Lax matrix we  adopted the convention $L= 2 L_\mathrm{Flaschka}$. This may look like an arbitrary choice, but is so to speak 
dictated by the structure of the Toda generalized free energy. This point was slightly overlooked in \cite{S19}, see also the arXiv version 4,
and for convenience we briefly repeat in Appendix \ref{sec8}.

As a next step we have to introduce a few standard items from generalized hydrodynamics. A generalized Gibbs state (GGE) of the Toda lattice is characterized by the pressure $P > 0$
and the chemical potential $V(w)$, see Appendix \ref{sec8}. $P$ is the thermodynamic dual to the stretch and $V(w)$ the dual to the 
collection of conserved charges, where instead of the discrete label $n$ a continuous label $w \in \mathbb{R}$ will be more convenient. To compute averages of the charges and their currents one starts from the free energy functional 
\begin{equation}\label{2.7}
\mathcal{F}(\varrho) =  \int _\mathbb{R}\mathrm{d}w \varrho(w) V(w)      - \int _\mathbb{R}\mathrm{d}w\int _\mathbb{R}\mathrm{d}w'   \log|w - w'|\varrho(w) \varrho(w') + 
\int _\mathbb{R}\mathrm{d}w \varrho(w) \log \varrho(w),
\end{equation} 
to be minimized under the constraint
\begin{equation}\label{2.8}
\varrho(w) \geq 0,\quad  \int _\mathbb{R}\mathrm{d}w\varrho(w) =P. 
\end{equation}
Introducing the Lagrange multiplier, $\mu$, the Euler-Lagrange equation for the minimizer, $\rho_\mu$, of $\mathcal{F}$ reads
\begin{equation}\label{2.9} 
  V(w) -  2 \int_\mathbb{R} \mathrm{d}w'  \log|w-w'| \rho_\mu(w') +\log \rho_\mu(w)  - \mu = 0.
 \end{equation}
Setting $\rho_\mu=  \mathrm{e}^{-\varepsilon}$ with quasi-energies $\varepsilon$, \eqref{2.9} turns into
the classical version of the TBA equation,
\begin{equation}\label{2.10} 
  \varepsilon(w) =  V(w) - \mu -  2 \int_\mathbb{R} \mathrm{d}w'  \log|w-w'|  \mathrm{e}^{-\varepsilon(w')}.
 \end{equation}

Let us define the integral operator 
\begin{equation}\label{2.11}
T\psi(w) = 2 \int_\mathbb{R} \mathrm{d}w' \log |w-w'| \psi(w'),\quad w \in \mathbb{R}.
\end{equation}
Then the  TBA equation is rewritten as 
\begin{equation}\label{2.12} 
\varepsilon (w)  = V(w)  -  \mu  - (T \mathrm{e}^{-\varepsilon})(w)
 \end{equation}
and one introduces the  dressing of a function $\psi$  by
\begin{equation}\label{2.13} 
\psi^\mathrm{dr} = \psi + T \rho_\mu \psi^\mathrm{dr},\quad \psi^\mathrm{dr} = \big(1 - T\rho_\mu\big)^{-1} \psi.
\end{equation}
On the right hand side $\rho_\mu$ is regarded as multiplication operator, $(\rho_\mu\psi)(w) = \rho_\mu(w)\psi(w) $.
The DOS (density of states) of the Lax matrix under GGE is given by $\nu\rho_\mathrm{p}$, such that
\begin{equation}\label{2.14} 
\partial_\mu \rho_\mu = \rho_\mathrm{p},  \quad \nu\langle\rho_\mathrm{p}\rangle = 1, \quad \nu = \langle r_0\rangle_{P,V},
 \end{equation}
where  $\langle r_0\rangle_{P,V}$ stands for the average with respect to the GGE state labeled by the parameters $P,V$, while $\langle f\rangle$ is merely a short hand for $\int_\mathbb{R} \mathrm{d}w f(w)$.

Differentiating TBA with respect to $\mu$ we conclude
\begin{equation}\label{2.15} 
\rho_\mathrm{p}= (1 - \rho_\mu T)^{-1} \rho_\mu = \rho_\mu(1 - T\rho_\mu)^{-1}[1] = \rho_\mu[1]^\mathrm{dr}.
 \end{equation}
 Here $[1]$ stands for the constant function, $\psi(w) = 1$, and similarly $[w^n]$ for the $n$-th power, $\psi(w) = w^n$.
 We also use $[w^n]^{\mathrm{dr}2} = ([w^n]^\mathrm{dr})^2$.
 A tracer pulse with bare velocity $v$ in fact travels with an effective velocity, which is determined through the integral equation 
 \begin{equation}\label{2.16} 
v^\mathrm{eff}(v) = v + (T\rho_\mathrm{p} v^\mathrm{eff})(v) - (T\partial_\mu\rho_\mathrm{p}(v))v^\mathrm{eff}(v).
 \end{equation}
 More concisely the effective velocity can be written as  
 \begin{equation}\label{2.17} 
v^\mathrm{eff} = \frac{ \,[w]^\mathrm{dr}}{\, [1]^\mathrm{dr}}. 
 \end{equation}
 
 With this input the GGE averaged charge densities are
\begin{equation}\label{2.18} 
\langle r_0\rangle_{P,V} = \nu, \quad \langle Q^{[n]}_0 \rangle_{P,V} =  \nu\langle \rho_\mathrm{p}w^n\rangle = q_n, \quad n = 1,2,...\,,
\end{equation}
with the corresponding average current densities
\begin{equation}\label{2.19} 
\langle J^{[0]}_0\rangle_{P,V} = -\langle Q^{[0]}_0\rangle_{P,V}, \quad \langle J^{[n]}_0 \rangle_{P,V} =  \langle (v^\mathrm{eff} - q_1)  \rho_\mathrm{p}w^n\rangle, \quad n = 1,2,...\,. \medskip
\end{equation}
The status of both identities is distinctly different. Under suitable conditions on the confining potential,  \eqref{2.18} is established in \cite{D19,S19},
while \eqref{2.19} should be viewed as a conjecture, however with strongly supporting numerical evidence \cite{BCS19}.\medskip\\
\textit{Remark on notations}. In \cite{D19} B. Doyon investigates the generalized hydrodynamics for the Toda lattice, both in  the fluid picture and the chain picture, the latter being adopted here.
Our notations differ only minimally. More precisely the definitions of $\nu$, $T$, $v^\mathrm{eff}$, and $\rho_\mathrm{p}$ are identical. $n$ in \cite{D19} becomes  $\rho_\mu$ here, 
because $n$ is used already otherwise.\medskip\,\,$\Box$

Note that through \eqref{2.16} $v^\mathrm{eff}$ becomes a functional of $\rho_\mathrm{p}$. Thus the Euler type equations for the Toda chain read
\begin{equation}\label{2.20} 
\partial_t \nu -\partial_x q_1 = 0,\qquad \partial_t(\nu \rho_\mathrm{p}) + \partial_x\big((v^\mathrm{eff} - q_1)\rho_\mathrm{p}\big) = 0.
\end{equation}
We claim that the nonlinear transformation $\rho_\mathrm{p} \mapsto \rho_\mu$,  defined by
\begin{equation}\label{2.21} 
\rho_\mu =  \rho_\mathrm{p}(1+ T\rho_\mathrm{p})^{-1},
\end{equation}
results in the normal form
\begin{equation}\label{2.22} 
\nu \partial_t \rho_\mu + (v^\mathrm{eff} - q_1)\partial_x \rho_\mu = 0.\medskip
\end{equation}
\textit{Proof}: Inserting in
\begin{equation}\label{2.23} 
 \rho_\mathrm{p}\partial_t\nu  + \nu\partial_t\rho_\mathrm{p} + \partial_x\big((v^\mathrm{eff} - q_1)\rho_\mathrm{p}\big) = 0,
\end{equation}
the continuity equation yields
\begin{equation}\label{2.24}
\nu\partial_t \rho_\mathrm{p} + \partial_x(v^\mathrm{eff} \rho_\mathrm{p})  - q_1 \partial_x \rho_\mathrm{p}= 0.
\end{equation}
Using this identity together with \eqref{2.21}, we obtain
\begin{eqnarray}\label{2.25}
&&\hspace{-20pt} \nu\partial_t \rho_\mu = \frac{\rho_\mu}{\rho_\mathrm{p}}\big(-\partial_x(v^\mathrm{eff}\rho_\mathrm{p}) +q_1\partial_x \rho_\mathrm{p}\big)
-\frac{\rho_\mu^2}{\rho_\mathrm{p}}\big(T(-\partial_x(v^\mathrm{eff}\rho_\mathrm{p}) + q_1\partial_x \rho_\mathrm{p})\big)\nonumber\\
&&\hspace{12pt} = - \frac{\rho_\mu}{\rho_\mathrm{p}}\partial_x\big((1 - \rho_\mu T)(v^\mathrm{eff}\rho_\mathrm{p})\big) + q_1 \partial_x\rho_\mu.
\end{eqnarray}
The effective velocity is solution of the integral equation
\begin{equation}\label{2.26}
v^\mathrm{eff} = v + T(v^\mathrm{eff}\rho_\mathrm{p}) - (T\rho_\mathrm{p})v^\mathrm{eff}.
\end{equation}
Hence
\begin{equation}\label{2.27}
\partial_x \big((1- \rho_\mu T)(v^\mathrm{eff}\rho_\mathrm{p})\big) =  -\frac{\rho_\mathrm{p}}{\rho_\mu}v^\mathrm{eff}\partial_x \rho_\mu.
\end{equation}
\noindent 
Adding all terms yields \eqref{2.22}. $\Box$
\section{GGE space-time covariance}\label{sec3}
\setcounter{equation}{0} 
\subsection{The static charge-charge and charge-current correlator}\label{sec3.1}
The space-time two-point function of the conserved charges under a prescribed
GGE state is defined by 
\begin{equation}\label{3.1}
S_{m,n}(j,t) = \big\langle Q_j^{[m]}(t)Q_0^{[n]}\big\rangle_{P,V} - \big\langle Q_0^{[m]}\big\rangle_{P,V}\langle Q_0^{[n]}\big\rangle_{P,V}, \quad m,n\geq 0.
\end{equation}
Here $t$ refers to the Toda time evolution with the convention $Q_j^{[n]}= Q_j^{[n]}(t=0)$. Through a Galilei transformation one can always achieve
$q_1 = 0$, which we adopt from now on.
An exact solution is of out reach and we focus on the hydrodynamic prediction for this correlator. 
On the Euler scale the initial randomness is merely propagated. The noise resulting from the dynamics becomes visible only
on the diffusive scale. Therefore  the linearized equation has to be solved with random initial data as deduced from the GGE. 
On the hydrodynamic scale they are delta-correlated in space with non-trivial charge correlations as determined through the static charge-charge covariance 
 \begin{equation}\label{3.2}
C_{m,n} = \big\langle Q^{[m]};Q^{[n]}\big\rangle_{P,V}, \quad C_{m,n} = C_{n,m}.
\end{equation}
Here we  use the shorthand 
\begin{equation}\label{3.3}
\big\langle Q;Q'\big\rangle_{P,V} = \sum_{j \in \mathbb{Z}} \big(\big\langle Q_0 Q'_j\big\rangle_{P,V} - \big\langle Q_0\big \rangle  \big\langle Q'_0\big\rangle_{P,V}\big).
\end{equation}

As to be explained in Sect. \ref{sec4}, for all $m,n\geq1$,
\begin{eqnarray}\label{3.4}
&&\hspace{-20pt} C_{0,0} =\nu^3 \langle \rho_\mathrm{p} [1]^\mathrm{dr} [1]^\mathrm{dr} \rangle,\nonumber\\[0.5ex]
&&\hspace{-20pt}C_{0,n}= C_{n,0}= -\nu^2  \langle \rho_\mathrm{p} [1]^\mathrm{dr} ([w^n]- q_n[1])^\mathrm{dr} \rangle,\nonumber\\[0.5ex]
&&\hspace{-20pt} C_{m,n}= \nu  \langle \rho_\mathrm{p} ([w^m]- q_m[1])^\mathrm{dr} ([w^n]- q_n[1])^\mathrm{dr} \rangle.
\end{eqnarray}
For later computations the basis consisting of moments is not so convenient. In addition the index $0$ plays a special role.  For these reasons we introduce the two
vector $(r,\phi)$ with $r\in\mathbb{R}$ and $\phi$ a real-valued function, formally $\phi(w) = \sum_{n=1}^\infty a_n w^n$ with some real coefficients $a_n$. The matrix then inherits a two-block structure.
We define 
\begin{equation}\label{3.5}
h = \nu\rho_\mathrm{p}, \quad \langle h \rangle = 1, \quad h \geq 0 ,
\end{equation}
and
\begin{equation}\label{3.6}
F\phi = (\phi - \langle h\phi\rangle)^\mathrm{dr} = \phi^\mathrm{dr} - [1]^\mathrm{dr}\langle h\phi\rangle = (1 - T\rho_\mu)^{-1}(\phi - \langle h\phi\rangle).
\end{equation}
Then
\begin{equation}\label{3.7}
C = 
\begin{pmatrix}
\nu^2 \langle h  [1]^{\mathrm{dr}2}\rangle &-\nu \big\langle F^\mathrm{T}h[1]^\mathrm{dr}\big|\\[0.5ex]
-\nu \big| F^\mathrm{T}h[1]^\mathrm{dr}\big\rangle& F^\mathrm{T}h F
\end{pmatrix},
\end{equation}
where $^\mathrm{T}$ stands for transpose and we freely use the Dirac notation for row and column vectors.
$F$ is a linear operator acting on the functions to the right. For example $F^\mathrm{T}h F\phi$ means first 
to compute $F\phi$, then multiply the result by the function $h$, and finally act with $F^\mathrm{T}$. Note that $F1 = 0$.

We also introduce the static charge-current correlator
\begin{equation}\label{3.8}
B_{m,n} = \big\langle Q^{[m]};J^{[n]}\big\rangle_{P,V}.
\end{equation}
On abstract grounds, see Sect. \ref{sec4}, the matrix $B$ is symmetric. In addition the column $m=0$ 
is already determined by $C$, since $J^{[0]} = -Q^{[0]}$. As to be discussed, one arrives at
\begin{eqnarray}\label{3.9}
&&\hspace{-20pt} B_{0,0} =  - \langle r;Q_1\rangle_{P,V} = \nu^2 \langle \rho_\mathrm{p} [1]^\mathrm{dr} [w]^\mathrm{dr}\rangle,\nonumber\\[0.5ex]
&&\hspace{-20pt}B_{0,n}= B_{n,0}= -\big\langle Q^{[n]};Q^{[1]}\big\rangle_{P,V} = -\nu \langle \varrho_\mathrm{p} [w]^\mathrm{dr} ([w^n]- q_n[1])^\mathrm{dr}\rangle,\nonumber\\[0.5ex]
&&\hspace{-20pt} B_{m,n} = B_{n,m} =  \langle \rho [w]^\mathrm{dr} ([w^m]- q_m[1])^\mathrm{dr} ([w^n]- q_n[1])^\mathrm{dr}\rangle.
\end{eqnarray}
Using $[w]^\mathrm{dr} = v^\mathrm{eff} [1]^\mathrm{dr} $, one obtains the more concise operator form
\begin{equation}\label{3.10}
B = 
\nu^{-1}\begin{pmatrix}
\nu^2 \langle h v^\mathrm{eff} [1]^{\mathrm{dr}2}\rangle &-\nu \big\langle F^\mathrm{T}h v^\mathrm{eff}[1]^\mathrm{dr}\big|\\[0.5ex]
-\nu \big| F^\mathrm{T}hv^\mathrm{eff}[1]^\mathrm{dr}\big\rangle& F^\mathrm{T}hv^\mathrm{eff} F
\end{pmatrix}.
\end{equation}
\subsection{Main result}\label{sec3.2}
The particular structure of the matrices $C,B$ allows for an educated guess of the full space-time correlators in the hydrodynamic approximation.
As can be seen from the following section, it will require actually some efforts to confirm this guess. 

Let us start with a generic example of $\kappa$ conservation laws linearized around a spatially homogeneous equilibrium state, which are governed by
\begin{equation}\label{3.11}
\partial_t u_n(x,t) + \sum_{m=1}^{\kappa} \mathsf{A}_{n,m} \partial_x u_m(x,t) = 0,\quad n = 1,...,\kappa.
\end{equation} 
Here $x\in\mathbb{R}$ is the continuum approximation for the lattice $\mathbb{Z}$. The $\kappa\times \kappa$ linearization matrix $\mathsf{A}$ is $x$-independent, 
generically not symmetric, but ensured to have a right and left system of eigenvectors, $|\psi_\alpha\rangle, \langle\tilde{\psi}_\alpha |, \alpha = 1,...,\kappa$, and has real eigenvalues $c_\alpha,\alpha = 1,...,\kappa$. 
\eqref{3.11} has to be solved with random initial conditions with mean zero and covariance 
\begin{equation}\label{3.12}
\mathbb{E}\big(u_m(x,0)u_n(0,0)\big) = \delta(x) \mathsf{C}_{m,n}, 
\end{equation} 
where to distinguish from other averages we use $\mathbb{E}(\cdot)$ for expectation over the initial noise. As before, $\mathsf{C}$ denotes the static correlator and $\delta(x)$ 
reflects the exponential decay of spatial correlations for the underlying microscopic model. Since the spatial part in \eqref{3.11} is merely a translation proportional to $t$, 
the space-time correlator is given by  
\begin{equation}\label{3.13}
\mathbb{E}(u_m(x,t)u_n(0,0)) = \sum_{\alpha = 1}^{\kappa} \delta(x-c_\alpha t) \big(|\psi_\alpha\rangle \langle\tilde{\psi}_\alpha |\mathsf{C}\big)_{m,n},
\end{equation} 
which has a simple interpretation: The $\alpha$-th peak travels with  velocity $c_\alpha$ and has a weight depending on $\mathsf{A}$ and $\mathsf{C}$. The static field-current correlator equals
\begin{equation}\label{3.14}
\mathsf{B} = \mathsf{A}\mathsf{C} = \tfrac{d}{dt} \mathrm{e}^{\mathsf{A}t}\mathsf{C}\big |_{t=0} = \sum_{\alpha = 1}^{\kappa} c_\alpha |\psi_\alpha\rangle \langle
\tilde{\psi}_\alpha | \mathsf{C}.
\end{equation} 
 
 Returning to the Toda chain, the charge-charge correlator $C$ has been computed in \eqref{3.7} and charge-current $B$ in \eqref{3.10}. To be in agreement with \eqref{3.14},
 the natural guess for the full solution reads
 \begin{equation}\label{3.15}
S(j,t) \simeq 
   \begin{pmatrix}
\nu^2 \langle h \,\delta(x - t\nu^{-1}v^\mathrm{eff}) [1]^{\mathrm{dr}2}\rangle &-\nu \big\langle F^\mathrm{T}h\,\delta(x - t\nu^{-1}v^\mathrm{eff})[1]^\mathrm{dr}\big|\\[0.5ex]
-\nu \big| F^\mathrm{T}h\,\delta(x - t\nu^{-1}v^\mathrm{eff})[1]^\mathrm{dr}\big\rangle& F^\mathrm{T}h\delta(x - t\nu^{-1}v^\mathrm{eff}) F
\end{pmatrix}.
\end{equation}
This expression is our \textbf{main result}, which is exact on the ballistic Euler scale. The right hand side for $t=0$ equals $\delta(x) C$ and 
differentiating with respect to $t$ at $t=0$ yields the required identity $B =AC$. Still to be shown is the exponential form, $\mathrm{e}^{At}C$, of the correlator, see Section \ref{sec4}.

Manifestly, the right hand side of Eq. \eqref{3.15} scales ballistically. The left hand side is the true correlator of the Toda chain which 
exhibits ballistic scaling only for sufficiently large $j,t$ both of the same order. This is the meaning  of $\simeq$, where one should identify $x$ with $j$ up to a suitable scale factor.
 
\eqref{3.15} is an operator identity. For specific correlations one has to compute the respective matrix elements.
As examples, we list the time-dependent stretch-stretch  and momentum-momentum correlation,  
\begin{eqnarray}\label{3.16}
&&S_{0,0}(j,t) \simeq \nu^2 \int_\mathbb{R}\mathrm{d}w  h(w) \delta(x- t\nu^{-1}v^\mathrm{eff}(w)) [1]^{\mathrm{dr}2}(w),\\
&&S_{1,1}(j,t) \simeq  \int_\mathbb{R}\mathrm{d}w  h(w) \delta(x- t\nu^{-1}v^\mathrm{eff}(w)) [v]^{\mathrm{dr}2}(w),
\end{eqnarray}
where in the second line $q_1 = 0$ has been used.
\section{Towards the exact solution}\label{sec4}
\setcounter{equation}{0} 
\subsection{The $C$- and $B$-matrix}\label{sec4.1}
We first collect a few identities, which will be used to compute the derivatives of the average charges and currents with respect to
the pressure $P$ and the chemical potential $V(w)$. The most basic identity is 
\begin{equation}\label{4.1}
 \partial_{\odot} \partial_\mu \rho_\mu= (1- \rho_\mu T)^{-1}\partial_\odot \rho_\mu(1- T\rho_\mu)^{-1},
\end{equation} 
which follows from expanding $\rho_\mathrm{p} = \partial_\mu \rho_\mu = (1- \rho_\mu T)^{-1} \rho_\mu$  in a power series.
Hence, setting $\odot = \mu$,
 \begin{equation}\label{4.2} 
\partial_\mu \langle \rho_\mathrm{p} \rangle= \langle(1- \rho_\mu T)^{-1}\partial_\mu \rho_\mu(1-  T\rho_\mu)^{-1} \rangle
  =\langle \rho_\mathrm{p}[1]^{\mathrm{dr} 2}\rangle.
\end{equation} 
Furthermore, for a general function $f$,
 \begin{equation}\label{4.3} 
\partial_P \langle \rho_\mathrm{p}f\rangle =  \partial_\mu\langle \rho_\mathrm{p}f\rangle \partial_P\mu = \nu\langle \rho_\mathrm{p}[1]^\mathrm{dr}[f]^\mathrm{dr}\rangle.
\end{equation} 
The variational derivative with respect to $V(w)$  is slightly more complicated,  since the constraint \eqref{2.8} has to be respected.
The corresponding directional derivative, oriented
along $g(w)$, is denoted by 
$\mathcal{D}_g  = \int \mathrm{d} w g(w) \delta/ \delta V(w)$. Then, for general functions $g,f,\tilde{f}$, one obtains
\begin{equation}\label{4.4} 
-\mathcal{D}_g \langle \tilde{f} (1- \rho_\mu T)^{-1} \rho_\mu  f\rangle =   \langle \rho_\mathrm{p} g^\mathrm{dr} \tilde{f}^\mathrm{dr}  f^\mathrm{dr}\rangle 
- \nu  \langle \rho_\mathrm{p}g^\mathrm{dr}  \rangle  \langle  \rho_\mathrm{p}  \tilde{f}^\mathrm{dr} f^\mathrm{dr}\rangle.
\end{equation} 

The average charges are
\begin{equation}\label{4.5} 
\big\langle r_0\big\rangle_{P,V} = \nu, \quad \big\langle Q^{[n]}_0 \big\rangle_{P,V} =  \nu\langle\rho_\mathrm{p}w^n\rangle.
\end{equation} 
By suitable linear combinations the powers $w^n$ are replaced by a general function $f(w)$. The covariance matrix $C$ is obtained by taking derivatives of
\eqref{4.5} with respect to $P$ and  $V$.

For $C_{0,0}$ we note
\begin{equation}\label{4.6}
-\partial_P \nu =  \nu^2 \partial_P  \langle \rho_\mathrm{p} \rangle = \nu^3 \langle \rho_\mathrm{p}[1]^{\mathrm{dr} 2}\rangle
\end{equation} 
and for $C_{0,n}, n \geq 1$ we use \eqref{4.3} to arrive at
\begin{eqnarray}\label{4.7}
&&\hspace{-40pt}- \partial_P\big( \nu\langle  \rho_\mathrm{p} f\rangle\big) = -(\partial_P\nu)\langle  \rho_\mathrm{p} f\rangle
- \nu \partial_P \langle  \rho_\mathrm{p} f\rangle
\nonumber\\
&&\hspace{32pt}=  \nu^3 \langle \rho_\mathrm{p}[1]^{\mathrm{dr} 2}\rangle\langle \rho_\mathrm{p} f\rangle - \nu^2 
\langle \rho_\mathrm{p}[1]^\mathrm{dr} [f]^\mathrm{dr}\rangle =-\nu \langle F^\mathrm{T}h[1]^\mathrm{dr}|f\rangle,
\end{eqnarray}
as claimed in \eqref{3.7}.
For $m,n\geq 1$ we use \eqref{4.4} for the special choice $\tilde{f} = 1$. Then
\begin{eqnarray}\label{4.8}
&&\hspace{-36pt}-\mathcal{D}_g(\nu\langle \rho_\mathrm{p} f \rangle) = \nu^2\big( \mathcal{D}_g\langle \rho_\mathrm{p}  \rangle\big) \langle \rho_\mathrm{p} f\rangle
-\nu\mathcal{D}_g \langle \rho_\mathrm{p} f\rangle = 
\nu^2\big(- \langle \rho_\mathrm{p} [1]^{\mathrm{dr} 2} g^\mathrm{dr} \rangle\nonumber\\ 
&&\hspace{46pt}
 + \,\nu \langle \rho_\mathrm{p} g \rangle\langle \rho_\mathrm{p} [1]^{\mathrm{dr} 2} \rangle\big) \langle \rho_\mathrm{p} f \rangle
- \nu \big( - \langle \rho_\mathrm{p} [1]^\mathrm{dr}g^\mathrm{dr}f^\mathrm{dr} \rangle + \nu \langle \rho_\mathrm{p} g \rangle \langle \rho_\mathrm{p} 
[1]^\mathrm{dr}f^\mathrm{dr} \rangle\big)\nonumber\\
&&\hspace{34pt} = \nu \langle \rho_\mathrm{p}
(g- [1] \nu  \langle \rho_\mathrm{p} g\rangle)^\mathrm{dr} (f- [1]\nu  \langle \rho_\mathrm{p}f \rangle)^\mathrm{dr}
\rangle,
\end{eqnarray}
in agreement with \eqref{3.7}.

The $B$-matrix is symmetric. Since this property is a useful control check, for conciseness, we repeat the argument 
in our specific context, see also \cite{S91}.  By stationarity $S_{m,n}(j,t) = S_{n,m}(-j,-t)$ and hence
\begin{equation}\label{4.9}
\sum_{j \in \mathbb{Z}} jS_{m,n}(j,t) = - \sum_{j \in \mathbb{Z}} jS_{n,m}(j,-t).
\end{equation}
Because of  the conservation law,
\begin{eqnarray}\label{4.10}
&&\hspace{-30pt}
 \frac{d}{dt}\sum_{j \in \mathbb{Z}} jS_{m,n}(j,t) =  \sum_{j \in \mathbb{Z}} j
\langle   J^{[m]}_{j-1}(t) -   J^{[m]}_{j}(t));Q^{[n]}_0(0)\rangle 
= \sum_{j \in \mathbb{Z}} \langle   J^{[m]}_j(t);Q^{[n]}_0(0)\rangle \nonumber\\
&&\hspace{59pt} =   \sum_{j \in \mathbb{Z}} \langle  J^{[m]}_0(0);Q^{[n]}_{-j}(-t) \rangle 
= \sum_{j \in \mathbb{Z}} \langle  J^{[m]}_0(0); Q^{[n]}_{j}(0) \rangle,
\end{eqnarray}
where we used stationarity  again and in the last step the conservation of $Q^{[n]}$.
Thus,
\begin{equation}\label{4.11}
\sum_{j \in \mathbb{Z}} jS_{m,n}(j,t) =  B_{m,n}t +\sum_{j \in \mathbb{Z}} jS_{m,n}(j,0) \,.
\end{equation}
Upon inserting in \eqref{4.9} one arrives at $B_{m,n}t = B_{n,m}t$.

Returning to Toda,  the average currents are
\begin{equation}\label{4.12}
\big\langle J^{[n]}_0 \big\rangle_{P,V} =   \langle (v^\mathrm{eff} - q_1) \rho_\mathrm{p}w^n\rangle =
 \langle w (1- \rho_\mu T)^{-1} \rho_\mu w^n\rangle -  \nu\langle  \rho_\mathrm{p}w\rangle \langle \rho_\mathrm{p}w^n\rangle.
\end{equation}
The second version will be more convenient for us. Note that while one can shift to $q_1 = 0$, when taking derivatives this term has to be kept.
As before the powers $w^n$ are replaced by a general function $f(w)$. Since $J^{[0]} = - Q^{[1]}$, the border matrix elements are already determined. As  control we still repeat with the result
\begin{eqnarray}\label{4.13}
&&\hspace{-30pt}-\partial_P \big( \langle w (1- \rho_\mu T)^{-1} \rho_\mu f \rangle - \nu \langle\rho_\mathrm{p}f\rangle \langle \rho_\mathrm{p}w\rangle\big)\nonumber\\
&&\hspace{20pt}=- \big(\nu   \langle  \rho_\mathrm{p}[w]^\mathrm{dr}f^\mathrm{dr}\rangle + \nu^3\langle \rho_\mathrm{p}[1]^{\mathrm{dr} 2}\rangle\langle\rho_\mathrm{p}f\rangle \langle \rho_\mathrm{p}w\rangle
 \nonumber\\
&&\hspace{32pt}-\nu^2\langle \rho_\mathrm{p}[1]^\mathrm{dr}f^\mathrm{dr}\rangle \langle \rho_\mathrm{p}w\rangle - \nu^2  \langle\rho_\mathrm{p}f\rangle \langle \rho_\mathrm{p}[w]^\mathrm{dr}[1]^\mathrm{dr}\rangle
\big).
\end{eqnarray}
The two middle terms vanish, since $q_1 = 0$, and the remainder agrees with \eqref{3.9}.
Finally, using \eqref{4.4} with $\tilde{f}(w) = w$, we turn to 
\begin{eqnarray}\label{4.14}
&&\hspace{-30pt}-\mathcal{D}_g \big( \langle w (1- \rho_\mu T)^{-1} \rho_\mu f \rangle - \nu \langle\rho_\mathrm{p}w\rangle \langle \rho_\mathrm{p}f\rangle\big)\nonumber\\
&& =  \langle \rho_\mathrm{p}( [w]^\mathrm{dr} - \nu\langle  \rho_\mathrm{p}w\rangle)(g - \nu \langle \rho_\mathrm{p} g\rangle [1])^\mathrm{dr}(f - \nu \langle \rho_\mathrm{p} f\rangle[1] )^\mathrm{dr} \rangle.
\end{eqnarray}
Since $q_1 = 0$, we have obtained the $(2,2)$ matrix element of  \eqref{3.10}. 
\subsection{Linearized transformation to normal modes}\label{sec4.2}
In \eqref{2.21} we introduced a nonlinear map which transforms the system of conservation laws into its quasi-linear version in such a way that the linearized operator $A$ is manifestly diagonal. One would expect this property to persist under linearization. In the $\nu,h$ variables the map is
\begin{equation}\label{4.15}
\rho_\mu = h(\nu + Th)^{-1}.
\end{equation} 
We linearize on both sides as $\rho_\mu +\epsilon g$, $\nu +\epsilon r$, $h +\epsilon \phi$, $\langle \phi \rangle = 0$. To first order in $\epsilon$
this yields the linear map   
$R: g \mapsto (r,\phi)$  given by  
\begin{equation}\label{4.16}
Rg = \nu
\begin{pmatrix} 
- \nu \langle g [1]^{\mathrm{dr}2}\rangle\\[0.5ex]
F^\mathrm{T}g [1]^\mathrm{dr}
\end{pmatrix},
\end{equation} 
where 
\begin{equation}\label{4.17}
F^{\mathrm{T}} \psi = (1 - \rho_\mu T)^{-1}\psi - h \langle [1]^\mathrm{dr} \psi\rangle.
\end{equation} 
Note that indeed $\langle F^{\mathrm{T}} \psi\rangle = 0$.  \eqref{4.15} can be inverted as  
\begin{equation}\label{4.18}
\nu = \langle \rho_\mathrm{p} \rangle^{-1},\quad h = \langle \rho_\mathrm{p} \rangle^{-1}  \rho_\mathrm{p}, \quad \rho_\mathrm{p} = (1- \rho_\mu T)^{-1}\rho_\mu,
\end{equation} 
thereby deriving, by a similar argument as before,
\begin{equation}\label{4.19}
R^{-1} 
\begin{pmatrix} 
r\\
\phi
\end{pmatrix}
= (\nu [1]^\mathrm{dr})^{-1}(- \rho_\mu r + (1-\rho_\mu T )\phi). 
\end{equation} 
Indeed one checks that
\begin{equation}\label{4.20}
RR^{-1} = 1,\quad R^{-1}R = 1,
\end{equation}
the first $1$ standing for the identity operator as a $2\times 2$ block matrix and the second $1$ for the identity operator in the space of scalar functions.

The next step is to write the $C, B$ matrices in the new basis.  Using
\begin{equation}\label{4.21}
(1-\rho_\mu T) F^\mathrm{T}\psi = \psi - \nu \rho_\mu \langle \psi[1]^{\mathrm{dr}}\rangle,
\end{equation}
one arrives at
\begin{eqnarray}\label{4.22}
&&\hspace{-47pt}R^{-1}CRg =  \nu R^{-1}
\begin{pmatrix}
\nu^2 \langle h  [1]^{\mathrm{dr}2}\rangle &-\nu \big\langle F^\mathrm{T}h[1]^\mathrm{dr}\big|\nonumber\\[0.5ex]
-\nu \big| F^\mathrm{T}h[1]^\mathrm{dr}\big\rangle& F^\mathrm{T}h F
\end{pmatrix}
\begin{pmatrix}
- \nu \langle g [1]^{\mathrm{dr}2}\rangle\\[0.5ex]
F^\mathrm{T}g [1]^\mathrm{dr}
\end{pmatrix}\\ [1ex]
&&\hspace{0pt}= ([1]^\mathrm{dr})^{-1}\big(\nu^3  \rho_\mu \langle h  [1]^{\mathrm{dr}2}\rangle  \langle g[1]^{\mathrm{dr}2}\rangle
+\nu \rho_\mu \langle (F^\mathrm{T}h[1]^\mathrm{dr})(F^\mathrm{T}g[1]^\mathrm{dr})\rangle\nonumber\\[0.5ex]
&&\hspace{30pt}+ \nu^2 (1-\rho_\mu T) F^\mathrm{T}h[1]^\mathrm{dr}\langle g[1]^{\mathrm{dr}2}\rangle
+ (1-\rho_\mu T) F^\mathrm{T}hFF^\mathrm{T}g[1]^\mathrm{dr}\big)\nonumber\\ [1ex]
&&\hspace{0pt} = \nu^2\big|([1]^\mathrm{dr})^{-2}h\big\rangle  \langle h  [1]^{\mathrm{dr}2}\rangle  \langle g[1]^{\mathrm{dr}2}\rangle
+ \big|([1]^\mathrm{dr})^{-2}h\big\rangle \langle (F^\mathrm{T}h[1]^\mathrm{dr})(F^\mathrm{T}g[1]^\mathrm{dr})\rangle\nonumber\\[0.5ex]
&&\hspace{30pt}+ \nu^2 \big|h\big\rangle\langle g[1]^{\mathrm{dr}2}  \rangle - \nu^2\big|([1]^\mathrm{dr})^{-2}h\big\rangle
 \langle h  [1]^{\mathrm{dr}2}\rangle  \langle g[1]^{\mathrm{dr}2}\rangle\nonumber\\[0.5ex]
&&\hspace{30pt}+\big|([1]^\mathrm{dr})^{-1} h FF^\mathrm{T}g[1]^\mathrm{dr} \big\rangle - \big|([1]^\mathrm{dr})^{-2} h\big\rangle \langle [1]^\mathrm{dr}hFF^\mathrm{T}g[1]^\mathrm{dr} 
\rangle.  
\end{eqnarray}
There are two cancellations which yield, as operators,
\begin{equation}\label{4.23}
R^{-1}CR =   \nu^2 \big| h \big\rangle \big\langle[1]^{\mathrm{dr}2}\big| + ([1]^\mathrm{dr})^{-1}h FF^\mathrm{T}[1]^\mathrm{dr}.
\end{equation}

Correspondingly for the $B$-matrix,
\begin{eqnarray}\label{4.24}
&&\hspace{-24pt}R^{-1}BRg =  R^{-1}
\begin{pmatrix}
\nu^2 \langle h v^\mathrm{eff} [1]^{\mathrm{dr}2}\rangle &-\nu \big\langle F^\mathrm{T}h v^\mathrm{eff}[1]^\mathrm{dr}\big|\nonumber\\[0.5ex]
-\nu \big| F^\mathrm{T}hv^\mathrm{eff}[1]^\mathrm{dr}\big\rangle& F^\mathrm{T}hv^\mathrm{eff} F
\end{pmatrix}
\begin{pmatrix}
- \nu \langle g [1]^{\mathrm{dr}2}\rangle\\[0.5ex]
F^\mathrm{T}g [1]^\mathrm{dr}
\end{pmatrix}\\ [1ex]
&&\hspace{14pt}= (\nu[1]^\mathrm{dr})^{-1}\big( \nu^3 \rho_\mu  \langle h  v^\mathrm{eff} [1]^{\mathrm{dr}2}\rangle  \langle g[1]^{\mathrm{dr}2}\rangle
+ \nu\rho_\mu \langle (F^\mathrm{T}h v^\mathrm{eff}
[1]^\mathrm{dr})(F^\mathrm{T}g[1]^\mathrm{dr})\rangle\nonumber\\[0.5ex]
&&\hspace{44pt}+ \nu^2 (1-\rho_\mu T) F^\mathrm{T}hv^\mathrm{eff}[1]^\mathrm{dr}\langle g[1]^{\mathrm{dr}2}\rangle
+ (1-\rho_\mu T) F^\mathrm{T}hv^\mathrm{eff}FF^\mathrm{T}g[1]^\mathrm{dr}\big)\nonumber\\ [1ex]
&&\hspace{14pt}= \nu^{-1}\big(\nu^2 \big|([1]^\mathrm{dr})^{-2}  h \big\rangle \langle h  v^\mathrm{eff} [1]^{\mathrm{dr}2}\rangle  \langle g[1]^{\mathrm{dr}2}\rangle
+ \big|([1]^\mathrm{dr})^{-2}h\big\rangle \langle (F^\mathrm{T}h v^\mathrm{eff}
[1]^\mathrm{dr})(F^\mathrm{T}g[1]^\mathrm{dr})\rangle\nonumber\\[0.5ex]
&&\hspace{44pt}+\nu^2\big| hv^\mathrm{eff}\big\rangle\langle g[1]^{\mathrm{dr}2}\rangle
- \nu^2 \big|([1]^\mathrm{dr})^{-2}h\big\rangle\langle hv^\mathrm{eff}[1]^{\mathrm{dr}2}\rangle\langle g[1]^{\mathrm{dr}2}\rangle
\nonumber \\[0.5ex]
&&\hspace{44pt}+ \big|([1]^\mathrm{dr})^{-1}hv^\mathrm{eff}FF^\mathrm{T}g[1]^\mathrm{dr} \big\rangle
-  \big|([1]^\mathrm{dr})^{-2}h\big\rangle \langle hv^\mathrm{eff}[1]^\mathrm{dr}FF^\mathrm{T}g[1]^\mathrm{dr}\rangle\big).
\end{eqnarray}
As before there are two cancellations yielding
\begin{equation}\label{4.25}
R^{-1}BR
 = \nu^{-1}\big(\nu^2\big| hv^\mathrm{eff}\big\rangle\big\langle [1]^{\mathrm{dr}2}\big|
   +  ([1]^\mathrm{dr})^{-1}hv^\mathrm{eff}FF^\mathrm{T}[1]^\mathrm{dr} \big)
 = \nu^{-1} v^\mathrm{eff} R^{-1}CR.  
\end{equation}
In the new basis, as anticipated, $R^{-1}AR$ is simply multiplication by $\nu^{-1} v^\mathrm{eff}$.

We conclude that 
\begin{equation}\label{4.26}
\mathrm{e}^{At}C =  R\mathrm{e}^{(v^\mathrm{eff}/\nu)t}R^{-1}C.
\end{equation}
Working out the algebra, one arrives at 
\begin{equation}\label{4.27}
\mathrm{e}^{At}C =  \begin{pmatrix}
\nu^2 \langle h\mathrm{e}^{(v^\mathrm{eff}/\nu)t}  [1]^{\mathrm{dr}2}\rangle &-\nu \big\langle F^\mathrm{T}h\mathrm{e}^{(v^\mathrm{eff}/\nu)t} [1]^\mathrm{dr}\big|\\[0.5ex]
-\nu \big| F^\mathrm{T}h\mathrm{e}^{(v^\mathrm{eff}/\nu)t} [1]^\mathrm{dr}\big\rangle& F^\mathrm{T}h\mathrm{e}^{(v^\mathrm{eff}/\nu)t } F
\end{pmatrix}.
\end{equation}
As explained in the beginning of Sect. \ref{sec3.2}, with this input one can write the solution to the hydrodynamic  equations \eqref{2.20} linearized relative to some precsribed GGE 
and with random initial conditions having covariance matrix $\delta(x)C$. The result is the right hand side of \eqref{3.15}. 

While not used, just for completeness the matrix $A$ is recorded as 
\begin{equation}\label{4.28}
A =  \nu^{-1}Rv_\mathrm{eff}R^{-1} = 
\begin{pmatrix}
\langle \rho_\mu v^\mathrm{eff} [1]^{\mathrm{dr}}\rangle &- \big\langle  v^\mathrm{eff}[1]^\mathrm{dr}(1 -\rho_\mu T)\big|\\[0.5ex]
-\nu^{-1} \big| F^\mathrm{T}\rho_\mu v^\mathrm{eff}\big\rangle& \nu^{-1}F^\mathrm{T}v^\mathrm{eff} (1 -\rho_\mu T)
\end{pmatrix}.
\end{equation}

\section{The $C$-matrix from Dyson's Brownian motion}\label{sec5}
\setcounter{equation}{0} 

The variational problem \eqref{2.7} is closely linked to Dyson's Brownian motion with confining potential $V$. To make this section self-contained, some background material is required.
The precise connection to Toda will be at the end of this section.

We consider the stochastic particle system on $\mathbb{R}$ governed by
\begin{equation}\label{5.1}
dx_j(t) = -V'(x_j(t))dt + \frac{1}{N}\sum_{i = 1,i\neq j}^N \frac{2\alpha}{x_j(t) - x_i(t)} dt + \sqrt{2} db_j(t), \quad j = 1,...,N, \quad\alpha \geq 0 , 
\end{equation}
with $\{b_j(t), j = 1,...,N\}$ a collection of independent standard Brownian motions.  This is Dyson's Brownian motion in an external  potential $V$. 
The interaction has strength $\alpha/N$, which corresponds to a  standard mean-field limit. (As to be discussed, the proper identification will  $\alpha = P$).
Let us introduce the empirical density 
 \begin{equation}\label{5.2}
\rho_N(x,t) =  \frac{1}{N} \sum_{j = 1}^N \delta(x_j(t)-x).
\end{equation} 
If at the initial time $t = 0$, $\rho_N(x,0)$ converges in the limit $N \to \infty$ to a non-random density $\rho(x,0)$, then such a limit will hold for any $t >0$ and the limit density satisfies the nonlinear Fokker-Planck equation
\begin{equation}\label{5.3}
\partial_t \rho(x,t) = \partial_x\big( V_\mathrm{eff}'(x,t)\rho(x,t) + \partial_x  \rho(x,t)\big),
\end{equation} 
where the effective potential is defined by 
\begin{equation}\label{5.4}
V_\mathrm{eff}(x,t) = V(x) - \alpha (T \rho)(x,t)
\end{equation} 
with $T$ is the linear operator defined in \eqref{2.11}. For this convergence result even a proof is available \cite{CL97}. If $V$ is suitably confining, then Eq. \eqref{5.3} has a unique stationary solution $\rho_\mathrm{s}$, which can also be characterized 
as the minimizer of
\begin{equation}\label{5.5}
\mathcal{F}^\alpha(\varrho) =  \int _\mathbb{R}\mathrm{d}w \varrho(w) V(w)      -\alpha \int _\mathbb{R}\mathrm{d}w\int _\mathbb{R}\mathrm{d}w'   \log|w - w'|\varrho(w) \varrho(w') + 
\int _\mathbb{R}\mathrm{d}w \varrho(w) \log \varrho(w)
\end{equation} 
under the constraints $\varrho (x) \geq 0$, $\int \mathrm{d}x \varrho (x) = 1$.

Let us now consider the stationary dynamics with $\rho(x,0) = \rho_\mathrm{s}(x)$ and study the fluctuations of the density. It is convenient to integrate against some smooth test function $f$.  Then the scaled density fluctuations are 
\begin{equation}\label{5.6}
\phi_N(f,t) = \frac{1}{\sqrt{N}} \sum_{j=1}^N \Big(f(x_j(t)) - \int _\mathbb{R}\mathrm{d}x   \rho_\mathrm{s}(x) f(x)\Big) = \int_\mathbb{R} \mathrm{d}x f(x) \phi(x,t).
\end{equation}
For them a standard central limit theorem holds, compare with \cite{I01}, 
 \begin{equation}\label{5.7}
\lim_{N\to \infty} \phi_N(f,t) = \phi(f,t), 
\end{equation} 
where  $\phi(x,t)$ is governed the linear Langevin equation
\begin{equation}\label{5.8}
\frac{d}{dt} \phi(x,t) = \partial_x D\phi(x,t) + \sqrt{2\rho_\mathrm{s}(x)}\xi(x,t).
\end{equation}  
Here $\xi(x,t)$ is normalized space-time white noise  and $\partial_x D $ the linearized evolution operator with
\begin{equation}\label{5.9}
D =  V'_\mathrm{eff} - \alpha \rho_\mathrm{s} T\partial_x.
\end{equation} 
$V'_\mathrm{eff}$ is defined as in \eqref{5.4} with $\rho(x,t)$  substituted by $\rho_\mathrm{s}(x)$. For the following argument $V'_\mathrm{eff}$
and $\rho_\mathrm{s}$ act as multiplication operators, while $\partial = \partial_x$ is the differentiation operator,
which commutes with $T$, $T\partial = \partial T$.

The stationary solution to Eq. \eqref{5.8} is a Gaussian measure, its covariance denoted by $C^\sharp$, which is determined by 
\begin{equation}\label{5.10}
\langle D^\mathrm{T}\partial f,C^\sharp g\rangle + \langle f,C^\sharp D^\mathrm{T}\partial g\rangle = 2\langle \partial f, \rho_\mathrm{s} \partial g\rangle,
\end{equation} 
where $\langle \cdot,\cdot \rangle$ denotes the standard $L^2$ inner product, $\langle f,g \rangle = \int\mathrm{d}x f(x) g(x)$.
We claim that as an operator
\begin{equation}\label{5.11}
C^\sharp = (1 - \alpha \rho_\mathrm{s} T)^{-1}\rho_\mathrm{s} - \frac{1}{\langle [1],(1 - \alpha \rho_\mathrm{s} T)^{-1}\rho_\mathrm{s}[1]\rangle} \big|(1 - \alpha \rho_\mathrm{s} T)^{-1}\rho_\mathrm{s}\big\rangle \big\langle (1 - \alpha\rho_\mathrm{s}T)^{-1}\rho_\mathrm{s}\big|.
\end{equation} 
The second term ensures that there are no fluctuations in the number of particles, $C^\sharp [1] = 0$.\medskip\\
\textit{Proof}: We consider only the left most term, the other one following by symmetry, and have to show that  
\begin{equation}\label{5.12}
\langle \partial f,D \rho_\mathrm{s} (1 -\alpha T \rho_\mathrm{s})^{-1}g\rangle = \langle \partial f, \rho_\mathrm{s} \partial g\rangle,
\end{equation} 
and, upon $g$ replacing by $(1 -\alpha T \rho_\mathrm{s})g$,
\begin{equation}\label{5.13}
\langle \partial f,D \rho_\mathrm{s} g\rangle = \langle \partial f, \rho_\mathrm{s} \partial (1 -\alpha T \rho_\mathrm{s})g\rangle.
\end{equation} 
Since $\rho_\mathrm{s}$ is stationary, we have 
\begin{equation}\label{5.14}
\big(V' -\alpha \rho_\mathrm{s} T\partial + \partial\big)\rho_\mathrm{s} = 0,
\end{equation} 
which when inserted in \eqref{5.13} leads to the condition 
\begin{equation}\label{5.15}
\rho_\mathrm{s}  \partial g -\alpha \rho_\mathrm{s} T\partial ( \rho_\mathrm{s}g) = \rho_\mathrm{s} \partial (1 -\alpha T \rho_\mathrm{s})g,
\end{equation} 
 It follows from using $T\partial = \partial T$, thereby confirming our claim. \medskip$\Box$

The see the connection to the $(2,2)$ matrix element of \eqref{3.7}, one recalls that because of linear ramp argument the covariance for the conserved charges is given by
$\partial_\alpha(\alpha C^\sharp)$, an expression depending only on $\alpha  \rho_\mathrm{s}$. Thus differentiating with respect $\alpha$ becomes identical to differentiating with respect to $P$.    
Using this observation in \eqref{5.11}, one arrives at the four terms defining the $(2,2)$ matrix element, thus providing an alternative derivation for this particular contribution. 
The stretch does not seem to be explicitly linked to the Lax matrix and for the $(1,2)$ matrix element of $C$ one has to rely on the thermodynamic reasoning.
\section{Hard rods viewed as lattice model}\label{sec6}
\setcounter{equation}{0} 
Hard rods are more naturally viewed as a one-dimensional fluid. To make the connection with the Toda chain, we discuss here the chain point of view. The rod length is denoted by $a$.
The $r_j$'s are changing linearly in time at constant $p_j$'s. At the instant when $r_j = a$ with incoming momenta $p_j ,p_{j+1}$ the momenta are simply interchanged, resulting in $\tfrac{d}{dt} r_j >0$. The Euler equations could be easily guessed. But in our context it is more instructive to follow the systematic
route outlined in Sect. \ref{sec2}. 
$\nu$ denotes the mean distance between hard rods, $\nu > a$, and $h(v)$ is the normalized velocity distribution. Thus $\rho_\mathrm{p} = \nu^{-1}h$ 
and $u = \langle h v \rangle$. The phase shift equals $-a$ and hence
\begin{equation}\label{6.1}
T = -a |1\rangle \langle 1|, \qquad (1 - a T\rho_\mu)^{-1} =  1 - (1 + a \langle \rho_\mu \rangle)^{-1} |1\rangle \langle a\rho_\mu|. 
\end{equation}
Since $\rho_\mathrm{p} = [1]^\mathrm{dr}\rho_\mu$, one arrives at
\begin{equation}\label{6.2}
\rho_\mu =  (\nu - a)^{-1}h 
\end{equation}
and, using \eqref{2.17},
\begin{equation}\label{6.3}
v^\mathrm{eff}  = (\nu - a)^{-1}(\nu v -a u).
\end{equation}
Thus the equations of GHD read
\begin{equation}\label{6.4}
\partial_t \nu - \partial_x u = 0,\quad \partial_t h +\partial_x \big((\nu - a)^{-1}(v-u)h\big) = 0
\end{equation}
with the normal mode transform 
\begin{equation}\label{6.5}
\partial_t \rho_\mu + (\nu - a)^{-1}(v-u) \partial_x\rho_\mu = 0.
\end{equation}
The factor $(\nu - a)^{-1}$ is the equilibrium density at contact.

The static covariance equals
\begin{equation}\label{6.6}
C = 
\begin{pmatrix}
(\nu - a)^2 &0\\[1ex]
0& h - |h\rangle\langle h|
\end{pmatrix}
\end{equation}
and the charge-current correlator, setting $u = 0$,
\begin{equation}\label{6.7}
B = 
\begin{pmatrix}
0 &- \langle hv|\\[1ex]
-| hv\rangle& (\nu- a)^{-1}\big(hv - | hv\rangle\langle h| - | h\rangle\langle hv|\big)
\end{pmatrix}.
\end{equation}
The linearized operator $A$ is explicit,
\begin{equation}\label{6.8}
A = 
\begin{pmatrix}
0 &- \langle v|\\[1ex]
-(\nu- a)^{-2}| vh\rangle& (\nu- a)^{-1}\big(v - | h\rangle\langle v| \big)
\end{pmatrix}.
\end{equation}
Linearizing \eqref{6.5}, as before, yields the similarity transform 
\begin{equation}\label{6.9}
Rg = (\nu- a)
\begin{pmatrix} 
- (\nu - a) \langle g \rangle\\
g - h\langle g\rangle
\end{pmatrix},\quad
R^{-1}
\begin{pmatrix}
r\\ \phi
\end{pmatrix}
= (\nu- a)^{-1} ( \phi - (\nu - a)^{-1}h r).
\end{equation} 
With this input
\begin{eqnarray}\label{6.10}
&&\hspace{ -8pt}\mathrm{e}^{At}C  = \\[0.5ex]
&&\hspace{ -12pt}\begin{pmatrix}
(\nu - a)^{2}\langle  \mathrm{e}^{v^\mathrm{eff}t}h\rangle &(\nu - a)\big(- \langle \mathrm{e}^{v^\mathrm{eff}t}h | +  \langle\mathrm{e}^{v^\mathrm{eff}t}h\rangle
\langle h|\big) \\[1ex]
(\nu - a)\big(-|\mathrm{e}^{v^\mathrm{eff}t}h\rangle + 
| h\rangle\langle\mathrm{e}^{v^\mathrm{eff}t}h  \rangle\big)& \mathrm{e}^{v^\mathrm{eff}t}h - |\mathrm{e}^{v^\mathrm{eff}t}h\rangle\langle h| - |h\rangle\langle \mathrm{e}^{v^\mathrm{eff}t}h|
+ \langle\mathrm{e}^{v^\mathrm{eff}t}h\rangle  |h\rangle \langle h|
\end{pmatrix}. \nonumber
 \end{eqnarray}
The limit $t \to 0$ and its first derivative yield $C$ and $B$, respectively.

\appendix
\section{Free energy}
\label{sec8}
\setcounter{equation}{0} 
We briefly explain how the Toda generalized free energy is linked to the variational problem \eqref{2.7}, more details being provided in \cite{S19}. For free boundary conditions, $(L_N)_{1,N} = 0$, and in the variables of \eqref{2.3}, the Toda partition function reads
\begin{eqnarray}\label{8.1}
&&\hspace{-29pt} Z_\mathrm{toda} =  \int_{\mathbb{R}^N} \prod_{j=1}^{N}  \mathrm{d}b_j \int_{(\mathbb{R}_+)^{(N-1)}}  \prod_{j=1}^{N-1} \mathrm{d}a_j \frac{2}{a_j}
(a_j)^{2P} \mathrm{e}^{-\mathrm{tr}[V(L_N)]}\nonumber\\
&& = \int_{\mathbb{R}^N} \prod_{j=1}^{N}  \mathrm{d}b_j \int_{(\mathbb{R}_+)^{(N-1)}}  \prod_{j=1}^{N-1} \mathrm{d}a_j \frac{2}{a_j}
(a_j)^{2P}   \mathrm{e}^{(\frac{1}{2}p_j^2 + (a_j)^2 ) }  \mathrm{e}^{-\mathrm{tr}[\tilde{V}(L_N)]},
\end{eqnarray}
where $\tfrac{1}{2}w^2 + \tilde{V}(w) = V(w)$. Here $P$ is the pressure and $V$ the chemical potential. More precisely, the $n$-th charge is controlled
by the chemical potential $\mu_n$. The grand-canonical weight is therefore the exponential of $\sum_{n=1}^\infty\mu_n Q^{[n]}= \sum_{n=1}^\infty \mu_n 
\mathrm{tr}[L^n]$.
Introducing $V(w) =  -\sum_{n=1}^\infty\mu_n w^n$ the exponent can be written more concisely as $-\mathrm{tr}[V(L_N)]$.
We compare $Z_\mathrm{toda}$ with the Dumitriu-Edelman partition function, 
\begin{eqnarray}\label{8.2}
&&\hspace{-30pt}Z_\mathrm{dued} = \int_{\mathbb{R}^N} \prod_{j=1}^{N}  \mathrm{d}b_j \int_{(\mathbb{R}_+)^{(N-1)}}  \prod_{j=1}^{N-1} \mathrm{d}a_j \frac{2}{a_j}
(a_j)^{2P(j/N)}   \mathrm{e}^{(\frac{1}{2}p_j^2 + (a_j)^2 ) }  \mathrm{e}^{-\mathrm{tr}[\tilde{V}(L_N)]}\\
&& =
D_N(P)\int_{\mathbb{R}^N} \prod_{j=1}^{N}\mathrm{d} \lambda_j 
\exp\Big[- \sum_{j=1}^N\big( \tfrac{1}{2}(\lambda_j)^2 + \tilde{V}(\lambda_j)\big)- P \frac{1}{N}\sum_{i,j=1,i \neq j}^N \log|\lambda_i - \lambda_j|\Big]. 
\nonumber\\
&& =
D_N(P)\int_{\mathbb{R}^N} \prod_{j=1}^{N}\mathrm{d} \lambda_j 
\exp\Big[- \sum_{j=1}^N\big( V(\lambda_j)\big)- P \frac{1}{N}\sum_{i,j=1,i \neq j}^N \log|\lambda_i - \lambda_j|\Big]. 
\end{eqnarray}
This identity follows from the Dumitriu-Edelman theorem \cite{DE02} with the choice $\beta = 2P/N$ and exploiting that $\mathrm{tr}[\tilde{V}(L_N)]$ depends only on the eigenvalues of $L_N$.  Only with the choice of the multiplicative factor as \eqref{2.3}, the potential  term on the left recombines 
into $V(\lambda_j)$. The prefactor is given by
\begin{equation}\label{8.3}
D_N(P) = \Gamma(P)^{-1}\Gamma(1+ \tfrac{P}{N})^N \prod_{j=1}^N\frac{\Gamma(\tfrac{j}{N})}{\Gamma(1+\tfrac{j}{N})}
\end{equation}
and 
\begin{equation}\label{8.4}
\lim_{N \to \infty} N^{-1} \log D_N(P) = -\log P +1.
\end{equation}

On the left of Eq. \eqref{8.2} we now use the slow ramp of the pressure and on the right the convergence to the mean field free energy of the log gas. Thus
\begin{equation}\label{8.5} 
 \int_0^1 \mathrm{d}u F_\mathrm{toda}(uP) =  \mathcal{F}^\mathrm{MF}_P(\rho^*) +\log P  -1
\end{equation}
with $\rho^*$ the minimizer of the mean field free energy functional. It is of advantage to absorb $P$ into $\varrho$ through
$P \mathcal{F}_P^\mathrm{MF}(P^{-1}\varrho)=  \mathcal{F}(\varrho) -P\log P$. Then 
 \begin{equation}\label{8.6} 
 F_\mathrm{toda}(P) =  \partial_P \mathcal{F}(\varrho^*(P)) - 1
 \end{equation}
with $\mathcal{F}$ as in \eqref{2.7}. Further properties are discussed in \cite{S19}. In particular
\begin{equation}\label{8.8} 
  F_\mathrm{toda}(P) =  \mu(P).
 \end{equation}


\end{document}